# Coherent effects in a thin film of metamaterial

Sergei O. Elyutin[*], Sergei S. Ozhenko, Andrei I. Maimistov

Moscow Engineering Physics Institute, 115409 Kashirskoe shosse 31, Moscow, Russia

**ABSTRACT**

The refraction is theoretically considered of ultimately short pulses at an interface of two dielectrics that contains a thin film of nonlinear metamaterial. For the model of metamaterial composed of nanoparticles and magnetic nanocircuits (split-ring resonators) the equations are obtained suitable for describing the coherent responses of such film. The numerical simulation demonstrates the emergence of oscillatory echo in inhomogeneous system of meta-atoms. It is supposed that the reported methods are applicable for investigation of thin metamaterial films.

## 1. INTRODUCTION

Quite a long time ago the ideas were stated[1,2] of the media where the directions of phase velocity and Poyting vector were opposite. The unusual properties were predicted of the waves propagating in such materials. These materials were called "left-handed material" (LHM). The LHMs possess the negative refraction index, when the real parts of permittivity and permeability are both negative in some frequency range. This is the reason for another name for such media – the negative refractive index (NRI) media. The existence of LHM was experimentally demonstrated firstly in microwave band[9-12]. The experimental evidences of LHM in optics were reported in[13-17]. The results of investigations and the analysis of perspective application of NRI materials are reviewed in[18,19].

The recent theoretical studies[1-7] have caused a great interest to the producing of NRI media. The nanotechnology development has led to the production of composite materials containing the inclusions of nanoparticles, nanowires, carbon nanotubes, nanomagnets and photonic crystals with metallic structure elements. To underline the artificial nature of such media they call them metamaterials[8]. Not all metamaterials possess negative refraction index, and not for all frequencies the refraction index is negative. However, this term is convenient when it is necessary to point to an object rather than on physical property.

The concrete constructions of metamaterials depend on both the elementary structural units (metaatoms) and a spatial layout of those atoms. Thus different sorts of metamaterials may have rather diverse electrodynamic features. In this connection it is expedient to engage the representation of homogeneous effective medium with certain optical characteristics (refractive index, extinction, permittivity etc.) while studying macroscopical properties of such media.

In most cases, the metamaterials are produced in film geometry. Films of meta-material often demonstrate rather substantial absorption of incident radiation. For experimental study of both linear and nolinear optical features of such media the methods of coherent spectroscopy[20-26] and nonlinear optics of thin films can be successfully applied[27]. For example, the measuring of bistability thresholds of nonlinear layer covering a thin film of meta-material allows determining the permittivity and permeability of the film[28,29].

In the present work, the refraction is considered of ultimately short pulse of electromagnetic radiation at the nonlinear interface of dielectric media. The excitation pulse represents mainly a pulson – a signal of electromagnetic field containing only several periods of carrier field, i.e. being very short in time and in spatial extension. The nonlinearity is due to a thin film of nonlinear material. If the thickness of the film is less then the length of a pulse, the pulse envelope of reflected and refractive waves are coupled to the incident wave by jump conditions. The evolution of polarization and magnetization is governed by differential equations, which can be formulated based on meta-material model. To describe the propagation of electromagnetic waves in artificial metamedium the phenomenological model of oscillatory polarization and magnetization was used[7,30,31]. This model well matches the metamaterial made of periodically arranged nanowires and nanocircuits (split ring resonators (SRRs))[32-36]. The generalization of this model, accounting the nonlinear properties of meta-material, was proposed in[37,38].

In this report, we used the analogous model[39], where the nonlinearity of plasma vibrations in nanoparticles immersed in dielectric matrix provides the nonlinear response of the film. The magnetic properties of the film are described by linear SRR oscillators. The dispersion in the parameters of nanoparticles and SSRs leads to inhomogeneous broadening of resonance absorption lines. Thus, a pair of ultimately short pulses incident on such film produce the coherent responses in the form of photon echo. By our knowledge this is a first report of observation of echo response in metamaterial excited by extremely short pulses.

---

[*] soelyutin@email.mephi.ru

By variation of the carrier wavelength it is possible to shift from the frequency band where meta-material is PRI to the field where it is NRI. That is one of the reasons to presume that photon echo effect could be a diagnostic tool to make conclusion on the optical properties of metamaterial.

## 2. THIN FILM OF METAMATERIAL AT THE INTERFACE OF LINEAR DIELECTRICS

Let a thin film, which can be polarized and magnetized by the electromagnetic field of incident wave, is inserted between two dielectric media at $X = 0$. A film may be made of the material of any nature: either linear or nonlinear, PRI or NRI, consisted of resonance or non-resonance atoms and molecules. In the current problem this film is the film of metamaterial.

The dielectric media are characterized by permittivity and permeability $\varepsilon_1$ and $\mu_1$ for $x < 0$, and $\varepsilon_2$ and $\mu_2$ for $x > 0$. Let for simplicity $\mu_1 = \mu_2 = 1$. In this case the magnetic induction of dielectrics can be identified with a strength of magnetic field in a space out of film. The $Z$ axe is chosen in the plane of interface. By virtue of the choice of the plane interface the Maxwell equations split into two independent sets describing the waves of TE-type: $\mathbf{E} = (0, E_y, 0)$, $\mathbf{H} = (H_x, 0, H_z)$, and TM-type $\mathbf{E} = (E_x, 0, E_z)$, $\mathbf{H} = (0, H_y, 0)$. It is assumed that thickness of the film $l_f$ is less than a spatial length of radiation pulse.

### 2.1. Jump conditions on a thin layer on interface

To describe the passing of light pulse through a thin film of nonlinear matter it is necessary to find the jump condition by both sides of this film. For this purpose, following[21,22], let us consider Maxwell equations

$$\mathrm{rot}\,\mathbf{E} = -\frac{1}{c}\frac{\partial \mathbf{B}}{\partial t}, \quad \mathrm{rot}\,\mathbf{H} = \frac{1}{c}\frac{\partial \mathbf{D}}{\partial t}, \tag{1}$$

where $\mathbf{B} = \mathbf{H} + 4\pi\mathbf{M}$, $\mathbf{D} = \mathbf{E} + 4\pi\mathbf{P}$.

Both, the magnetization and the polarization in these formulae contain homogeneous part associated with dielectric media and singular (proportional to delta-function) part, associated with thin film. For the sake of definiteness we will focus on the case of TE wave.

Then equations (1) yields:

$$\frac{\partial E_y}{\partial z} = \frac{1}{c}\frac{\partial B_x}{\partial t}, \quad \frac{\partial E_y}{\partial x} = -\frac{1}{c}\frac{\partial B_z}{\partial t}, \quad \frac{\partial H_x}{\partial z} - \frac{\partial H_z}{\partial x} = \frac{1}{c}\frac{\partial D_y}{\partial t}. \tag{2}$$

If to integrate the second equation in (2) over $x$

$$\int_{-\delta}^{+\delta} \frac{\partial E_y}{\partial x} dx = -\frac{1}{c}\frac{\partial}{\partial t}\int_{-\delta}^{+\delta} B_z dx,$$

then with the account of $B_z = B_z(z, x \neq 0) + 4\pi M_z^{(s)}\delta(x)$ ($M_z^{(s)}$ is the surface magnetization) one can obtain

$$E_y(+\delta) - E_y(-\delta) = -\frac{4\pi}{c}\frac{\partial}{\partial t}M_z^{(s)} - \frac{2\delta}{c}\frac{\partial}{\partial t}B_z(\delta/2) + O(\delta^2).$$

Under $\delta$ tending to zero this yields:

$$E_y(0-) - E_y(0+) = \frac{4\pi}{c}\frac{\partial}{\partial t}M_z^{(s)}. \tag{3.1}$$

The integration over $x$ in the third equation of (2)

$$\int_{-\delta}^{+\delta} \frac{\partial H_x}{\partial z} dx - \int_{-\delta}^{+\delta} \frac{\partial H_z}{\partial x} dx = \frac{1}{c}\frac{\partial}{\partial t}\int_{-\delta}^{+\delta} D_y dx,$$

with the account of $D_y = D_y(z, x \neq 0) + 4\pi P_y^{(s)}\delta(x)$ ($P_y^{(s)}$ is the surface polarization), yields

$$-H_z(+\delta) + H_z(-\delta) + 2\delta \frac{\partial H_x(\delta)}{\partial z} = \frac{4\pi}{c}\frac{\partial}{\partial t}P_y^{(s)} + \frac{2\delta}{c}\frac{\partial}{\partial t}D_y(\delta/2) + O(\delta^2).$$

The following comes from above relationships, when $\delta$ goes to zero:

$$H_z(0-) - H_z(0+) = +\frac{4\pi}{c}\frac{\partial}{\partial t}P_y^{(s)} \tag{3.2}$$

Expressions (3.1) and (3.2) show, that due to permeability and permittivity of a thin film, the strength of electric and magnetic fields are different by opposite sides of the film. These conclusions are valid for both the continuous waves and the ultra-short or the ultimately short pulses of electromagnetic radiation. The relationships (3) are the generalization of the corresponded formulae in[20-27].

For the TE wave the jump conditions (3.1) and (3.2) can be utilized to derive the relationships between the strength of electromagnetic fields in incident, reflected and refracted pulses. It is assumed for simplicity that the dielectric media by both sides of the film are dispersionless ones. Then the solutions of Maxwell equations can be written in the following form:

$$E_y(x,t) = \begin{cases} E_{in}(t - x/V_1) + E_{ref}(t + x/V_2), & x < 0 \\ E_{tr}(t - x/V_2), & x > 0 \end{cases} \tag{4}$$

Here $V_1$ is a group velocity of solitary wave (pulse) propagation in a medium at $x < 0$, $V_2$ is a group velocity at $x > 0$. Due to the absence of dispersion in these media, the group velocities are the parameters which feature the media. The jump condition (3.1) provides the relationship:

$$E_{in}(t) + E_{ref}(t) - E_{tr}(t) = \frac{4\pi}{c}\frac{\partial}{\partial t}M_z^{(s)}. \tag{5}$$

It follows from Maxwell equations for planar symmetry that

$$\frac{\partial E_y}{\partial x} = -\frac{1}{c}\frac{\partial H_z}{\partial t}.$$

With the account of this expression equation (3.2) yields:

$$\frac{1}{c}\frac{\partial}{\partial t}H_z(x=0-) - \frac{1}{c}\frac{\partial}{\partial t}H_z(x=0+) = \frac{4\pi}{c^2}\frac{\partial^2}{\partial t^2}P_y^{(s)},$$

or

$$\frac{\partial}{\partial x}E_y(x=0-) - \frac{\partial}{\partial x}E_y(x=0+) = -\frac{4\pi}{c^2}\frac{\partial^2}{\partial t^2}P_y^{(s)}. \tag{6}$$

Making use of (2) and bearing in mind that the introduction of the retarded time $t' = t \pm x/V_{1,2}$ and $x = x'$ specifies the derivations as following

$$\frac{\partial}{\partial t} = \frac{\partial}{\partial t'} \text{ and } \frac{\partial}{\partial x} = \frac{\partial}{\partial x'} \pm \frac{1}{V_{1,2}}\frac{\partial}{\partial t'},$$

equation (6) can be then re-written as

$$\frac{\partial}{\partial t'}\left(E_{ref} - E_{in} + \frac{V_1}{V_2} E_{tr}\right) = -\frac{4\pi V_1}{c^2} \frac{\partial^2}{\partial t'^2} P_y^{(s)}(t'). \tag{7}$$

Equation (7) can be integrated over time, thus providing the relationship

$$E_{ref} - E_{in} + \frac{V_1}{V_2} E_{tr} = -\frac{4\pi V_1}{c^2} \frac{\partial}{\partial t'} P_y^{(s)}(t'), \tag{8}$$

The refracted $E_{tr}(t')$ and reflected $E_{ref}(t')$ waves can be obtained from (5) and (8) in order to write down the relationship generalizing the Fresnel formula to the case of anharmonic waves.

$$E_{tr}(t') = \frac{2V_2}{V_1+V_2} E_{in}(t') - \frac{4\pi V_2}{c(V_1+V_2)}\left(\frac{\partial}{\partial t'} M_z^{(s)}(t') + \frac{V_1}{c}\frac{\partial}{\partial t'} P_z^{(s)}(t')\right) \tag{9.1}$$

and

$$E_{ref}(t') = \frac{V_2-V_1}{V_1+V_2} E_{in}(t') + \frac{4\pi V_1}{c(V_1+V_2)}\left(\frac{\partial}{\partial t} M_z^{(s)}(t') - \frac{V_2}{c}\frac{\partial}{\partial t'} P_y^{(s)}(t')\right). \tag{9.2}$$

In order to obtain polarization and magnetization it is necessary to define the field inside the film. It was demonstrated in[40] by means of Green-function technique that these values can be redefined in a symmetric way:

$$E(x=0) = E(x=0-) + (1/2)[E(x=0+) - E(x=0-)] = (1/2)[E(x=0+) + E(x=0-)]. \tag{10.1}$$

Analogously,

$$H_{y,z}(x=0) = H_{y,z}(x=0-) + (1/2)[H_{y,z}(x=0+) - H_{y,z}(x=0-)] = (1/2)[H_{y,z}(x=0+) + H_{y,z}(x=0-)]. \tag{10.2}$$

If the film is a non-magnetic one a known result comes out from (10.1) for the case of TE wave[21, 22]:

$$E_y(x=0) = E_y(x=0+) = E_y(x=0-).$$

The field strength inside the film follows from (10):

$$E = [E_{tr}(t') + E_{in}(t') + E_{ref}(t')]/2, \tag{11.1}$$

$$H(t) = [(c/V_2)E_{tr}(t') + (c/V_1)(E_{in}(t') - E_{ref}(t'))]/2 = \left\{\sqrt{\varepsilon_2} E_{tr}(t') + \sqrt{\varepsilon_1} E_{in}(t') - \sqrt{\varepsilon_1} E_{ref}(t')\right\}/2. \tag{11.2}$$

where it was used that the media beyond the film are non-dispersive and phase velocity coincides with the group one. As an example the inner field for TE wave can be obtained from (9):

$$E(t', x=0) = \frac{2V_2}{V_1+V_2} E_{in}(t') + \frac{2\pi(V_1-V_2)}{c(V_1+V_2)} \frac{\partial}{\partial t'} M_z^{(s)}(t') - \frac{4\pi V_1 V_2}{c^2(V_1+V_2)} \frac{\partial}{\partial t'} P_y^{(s)}(t'),$$

$$H(t', x=0) = \frac{2c}{V_1+V_2} E_{in}(t') - \frac{2\pi(V_1-V_2)}{(V_1+V_2)} \frac{\partial}{\partial t'} P_y^{(s)}(t') - \frac{4\pi c}{(V_1+V_2)} \frac{\partial}{\partial t'} M_z^{(s)}(t').$$

If the media by both sides of the film are identical, these expressions simplify

$$E(t', x=0) = E_{in}(t') - \frac{4\pi}{c\sqrt{\varepsilon}} \frac{\partial}{\partial t'} P_y^{(s)}(t'), \quad H(t', x=0) = \sqrt{\varepsilon} E_{in}(t') - 2\pi\sqrt{\varepsilon} \frac{\partial}{\partial t'} M_z^{(s)}(t').$$

## 2.2. A thin film material

To obtain the polarization and magnetization of the layer material it is necessary to select the model describing this material. Oftenly[41,42] in the model of metamaterial the dielectric properties are accounted by means of Lorenz oscillators for plasma vibrations, but magnetic features are described by the ensemble of circuits – SRRs[32-36]. As nanoparticles and nanocircuits are the elementary objects of artificial medium, they call them the metaatoms for brevity.

The simplest generalization of this model can be achieved by an accounting the anharmonicity of the electrical oscillations in nanoparticles[39] or the insertion of nonlinear capacity in a circuit[37,38]. Let us use the following equations[39] for the bulk polarization and magnetization of thin film material per one meta-atom

$$\frac{d^2 P}{dt'^2} + \Omega_d^2 P + \Gamma_e \frac{dP}{dt'} + \alpha P^3 = \left(\Omega_p^2 / 4\pi\right) E, \quad (12.1)$$

$$\frac{d^2 M}{dt'^2} + \Omega_T^2 M + \Gamma_m \frac{dM}{dt'} = -\left(\beta_m / 4\pi\right) \frac{d^2 H}{dt'^2}, \quad (12.2)$$

where $\Omega_p = \sqrt{4\pi\eta e^2 m_e^{-1} n_e}$ is an effective plasma frequency, $n_e$ is conduction electron volume density, $\eta$ is the metal-filling factor – the fraction of the composite occupied by metal[39], $\Omega_T$ is the Thomson frequency for nanocircuit, factor $\beta_m$ is determined by the geometry of the nanostructures $\beta_m \sim \eta^{2/3}$, nonlinearity coefficient can be estimated as following[39,43] $\alpha = \left(m_e a \Omega_p^2 / \hbar\right)^2$. Here $a$, $e$ and $m_e$ are the radius of the nanoparticle, the electron charge and the rest mass, respectively. The losses due to the ohmic resistance in circuits and the damping of plasmonic oscillations are accounted by parameters $\Gamma_m$ and $\Gamma_e$. $\Omega_p \approx 2\pi \cdot 27,5$ GHz for Cu. In the case when anharmonicity constatnt $\kappa$ is equal to zero, the system of equations (12) coincide with one used in[30,31] to investigate propagation and refraction of electromagnetic waves in NRI media. If the parameters of meta-atoms differ, the polarization and magnetization (12) correspond to an individual meta-atom from the ensemble. The total polarization (and magnetization) in (9) is the result of averaging over this ensemble.

## 2.3. Basic equations of the model

Finally, the whole problem looks like that

$$e_{tr} = F_0 e_{in} - g \frac{d}{d\tau} \left(\langle p_x \rangle_d + \langle m_x \rangle_T\right) \quad (13.1)$$

$$e_{ref} = R_0 e_{in} - g \frac{d}{d\tau} \left(\langle p_x \rangle_d - n_{12} \langle m_x \rangle_T\right) \quad (13.2)$$

$$\frac{d^2 p_x}{d\tau^2} + \omega_d^2(x) p_x + \gamma_p \frac{dp_x}{d\tau} + \kappa p_x^3 = \tfrac{1}{2}\left(e_{tr} + e_{in} + e_{ref}\right) \quad (13.3)$$

$$\frac{d^2 m_x}{d\tau^2} + \omega_T^2(x) m_x + \gamma_m \frac{dm_x}{d\tau} = -\frac{\beta_m n_1^2}{2} \frac{d^2}{d\tau^2}\left(e_{tr} + e_{in} + e_{ref}\right) \quad (13.4)$$

Equations (13) are written in dimensionless variables:

$$\tau = t' \Omega_p, \ e_{tr} = E_{tr}/E_0, \ e_{ref} = E_{ref}/E_0, \ p_x = 4\pi P/E_0, \ m_x = 4\pi n_1 M/E_0, \ \kappa = \alpha \left(E_0/4\pi\Omega_p\right)^2,$$

$$\gamma_{p,m} = \Gamma_{p,m}/\Omega_p, \ n_{12} = n_2/n_1, \ g = 2\pi \left(l_f/\lambda_p\right)/(n_1 + n_2), \ \lambda_p = 2\pi c/\Omega_p.$$

$E_0$ is a charachteristic value of the field strength in the problem. It could be chosen as the amplitude of the incident pulse.

In (13) the angel brackets $\langle \ \rangle_d$, $\langle \ \rangle_T$ denote correspondingly the averaging over the ensembles of nanoparticles and nanocircuits with generally different statistical laws. The normalized dimensional quantization frequency $\omega_d$ and circuit frequency $\omega_T$ depends on the parameter of dimensional inhomogenity $x$ as individual functions $\omega_d(x) = \bar{\omega}_d d(x)$ and $\omega_T(x) = \bar{\omega}_T T(x)$. Parameters $\bar{\omega}_d$, $\bar{\omega}_T$ are the characteristic normalized frequencies of the above mentioned distributions. A specific form of functions $d(x)$, $T(x)$ may vary as it depends on the metamaterial arrangement proposed for example in[9,16,32,35,37]. Generally speaking, both functions $d(x)$ and $T(x)$ drop with the increase of the dimension of radiating area.

The coefficients $F_0$ and $R_0$ in (13) are

$$F_0 = n_{12}^{-\frac{1}{2}} \mathcal{F}_0^{\frac{1}{2}} = \frac{2}{1+n_{12}}, \quad R_0 = sgn(1-n_{12})\mathcal{R}_0^{\frac{1}{2}},$$

where

$$\mathcal{F}_0 = \frac{4n_{12}}{(1+n_{12})^2}, \quad \mathcal{R}_0 = \left(\frac{1-n_{12}}{1+n_{12}}\right)^2, \quad \mathcal{F}_0 + \mathcal{R}_0 = 1$$

are the Fresnel coefficients of transmission and reflection on the interface of two transparent media.

The quantities, which are adequate for pulsed radiation, are the energy transmission and reflection coefficients:

$$F = \frac{n_{12}\int_0^\infty e_{tr}(\tau')^2 d\tau'}{\int_0^\infty e_{in}(\tau')^2 d\tau'}, \quad R = \frac{\int_0^\infty e_{ref}(\tau')^2 d\tau'}{\int_0^\infty e_{in}(\tau')^2 d\tau'}. \tag{14}$$

### 2.4. The incident pulses

In the frames of current consideration, the incident field constitutes the ultimately short pulses (USP). The USP's called uniplolar pulses or videopulses do not possess carrier at all. The electromagnetic field of such pulses can be presented by the Gaussian shape:

$$E(x,t) = E_0 \exp\{-(t \pm x/c)^2 / 2t_p^2\}.$$

The spectrum of such pulse $E(0,\omega) = E_0 t_p \exp\{-t_p^2 \Omega^2 / 2\}$ is localized in the vicinity of $\Omega = 0$ with the full width on half maximum (FWHM) $\Delta\Omega = 2\sqrt{2\ln 2}\, t_p^{-1}$.

The other sort of USP is a pulse of quasiharmonic radiation, squeezed by special optical means down to the duration of about a period of electromagnetic field vibration in initial signal. This shape can be presented as

$$E(x,t) = E_0 \exp\{-t^2/2t_p^2\}\exp\{-i\Omega_0 t + ik_0 x\} + c.c.$$

where $\Omega_0$ is the pulsation frequency, $k_0 = \Omega_0/c$ and the condition $t_p\Omega_0 \sim 1$ (more precisely $t_p\Omega_0 \approx \pi^2/2$) satisfied. The spectrum of pulson is expressed by the formula:

$$E(x=0,t) = E_0 t_p \exp\{-t_p^2(\Omega-\Omega_0)^2/2\} + E_0 t_p \exp\{-t_p^2(\Omega+\Omega_0)^2/2\}.$$

Its FWHM is the same as for videopulse, but the whole spectrum is shifted on $\Omega_0$. By varying $\Omega_0$ one can drive the spectrum of pulson from a frequency band, where metamaterial is characterized by a positive refraction index, to the area of negative refraction. In particular, the transmission $F$ and reflection $R$ coefficients can be now considered as the function of $\omega_0 = \Omega_0/\Omega_p$.

The effective permittivity and effective permeability of bulk sample, which meet the model of metamaterial (12) in linear approximation, are given by the expressions:

$$\varepsilon(\omega) = \left(1 + \frac{\omega_p^2}{\bar{\omega}_d^2 - \omega^2}\right) = \frac{\omega_a^2 - \omega^2}{\bar{\omega}_d^2 - \omega^2}, \quad \mu(\omega) = \left(1 + \frac{\beta_m \omega^2}{\bar{\omega}_T^2 - \omega^2}\right) = (1 - \beta_m)\frac{\omega_b^2 - \omega^2}{\bar{\omega}_T^2 - \omega^2}, \quad (15)$$

where $\omega_a^2 = 1 + \bar{\omega}_d^2$, $\omega_b^2 = \bar{\omega}_T^2 (1-\beta_m)^{-1}$, $\omega_j = \Omega_j/\Omega_p$.

For the square of refraction index we have

$$n^2 = \varepsilon(\omega)\mu(\omega) = (1-\beta_m)\frac{(\omega_b^2 - \omega^2)(\omega_a^2 - \omega^2)}{(\bar{\omega}_T^2 - \omega^2)(\bar{\omega}_d^2 - \omega^2)} \quad (16)$$

### 3. NUMERICAL RESULTS

System of equations (13) was solved numerically by the iteration procedure of prediction and correction with a desired accuracy. System (13) contains a certain complexity as the polarization and magnetization for individual mataatom (13.3,4) are determined by the fields which in their turn depend on polarization and magnetization averaged over the

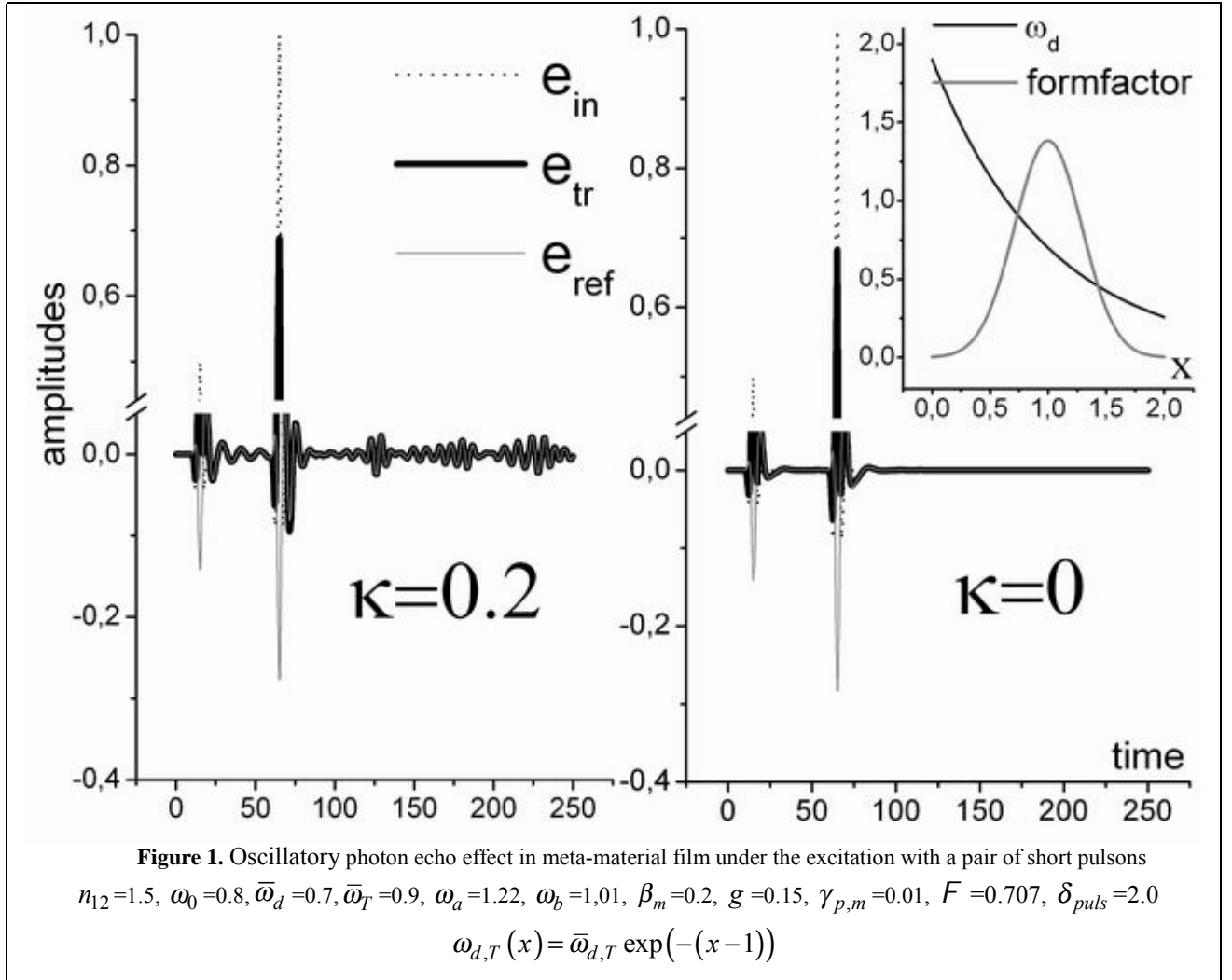

**Figure 1.** Oscillatory photon echo effect in meta-material film under the excitation with a pair of short pulsons
$n_{12}$ =1.5, $\omega_0$ =0.8, $\bar{\omega}_d$ =0.7, $\bar{\omega}_T$ =0.9, $\omega_a$ =1.22, $\omega_b$ =1,01, $\beta_m$ =0.2, $g$ =0.15, $\gamma_{p,m}$ =0.01, $F$ =0.707, $\delta_{puls}$ =2.0

$$\omega_{d,T}(x) = \bar{\omega}_{d,T} \exp(-(x-1))$$

inhomogeneous ensemble of metaatoms. That makes the problem self-consistent. This obstacle can be overcome with the additional iteration process involving transmitted and reflected field. The starting approximation for the fields could be their classical Fresnel values for empty interface. The iteration process over fields ends on reaching the assigned accuracy.

It should be emphasized that the nature of the observed response is the oscillatory echo[44] as it forms on the ensemble of nonlinear oscillators represented by equations (13.3) for plasmonic vibrations. In fact, due to the self-consistent form of (13) the magnetodipole subsitem also takes part in echo formation.

The oscillatory echo appears in the form of eqidistant train of signals at the moments after double excitation of the nonlinear medium by short pulses (fig.1 left panel). The plots in figure 1 demonstrate, that oscillatory echo effect completely vanishes if anharmonicity constatnt $\kappa$ is zero (fig.1 right panel). The Gauss form-factor of inhomogeneous ensemble of metaatoms and the dependence of the dimensional quantization frequency $\omega_d$ on the dimensional parameter $x$ are depicted on the inset.

The excitation of metamedia with pulson provides the opportunity to shift the spectrum band of pulson over the frequency domain by varying the pulsation frequency $\omega_0$. But, in order to operate with a distinct spectral band of pulson, one needs to satisfy the condition $t_p\Omega_0 \approx \pi^2/2$, rather then simply $t_p\Omega_0 \sim 1$. That means the shift of the frequency must be accompanied by the variation of pulse total duration.

Our preliminary calculations show that the oscillatory echo effect in the ensemble of metaatoms is stronger when the pulsation frequency falls into the NRI band. This case is depicted in figure 2, where echo is generated by a multiperiod pulson. In the inset there is a frequency layout of metamaterial parameters were dots point to the zeroes and poles of $\varepsilon$, $\mu$ and $n_{film}$ (15,16). It is seen that in this particular case the pulsation frequency get into NRI area.

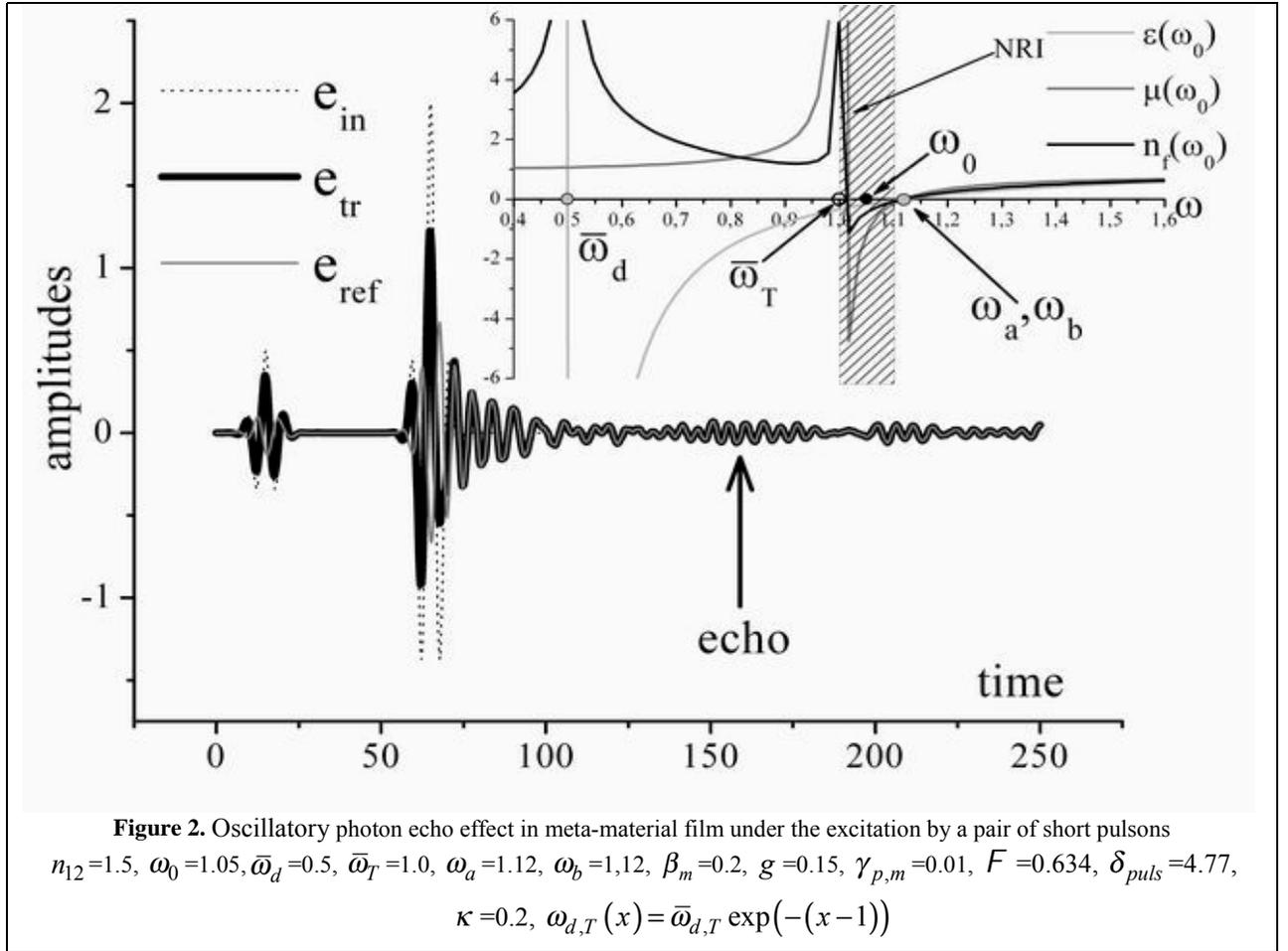

**Figure 2.** Oscillatory photon echo effect in meta-material film under the excitation by a pair of short pulsons
$n_{12}$=1.5, $\omega_0$=1.05, $\bar{\omega}_d$=0.5, $\bar{\omega}_T$=1.0, $\omega_a$=1.12, $\omega_b$=1,12, $\beta_m$=0.2, $g$=0.15, $\gamma_{p,m}$=0.01, $F$=0.634, $\delta_{puls}$=4.77, $\kappa$=0.2, $\omega_{d,T}(x) = \bar{\omega}_{d,T}\exp(-(x-1))$

Figure 3 displays the temporal shape of the transmitted and reflected pulses when the frequency of pulsation $\omega_0$ changes. In the vicinity of electrodipole and partially magnetodipole resonance the well recognized nutational motion of the field appears in the after-pulse moments of time. That clearly manifests the coherent character of pulse interaction with mataatoms of the film.

In the inset the dependences of transmission ant reflection coefficients vs pulson frequency demonstrate the effectiveness of USP interaction with metamedia at resonances contained in the model of the film.

## 5. CONCLUSION

In conclusion, we considered the refraction of ultimately short electromagnetic pulse in the form of pulson at a thin layer of metamaterial placed at the interface between two dielectric media. Under such excitation and for such geometry of a sample the conventional permittivity and the permeability cannot be introduced. But still, effective permeability could be consider, and these parameters can change sign depending on the frequency of shape modulation in pulson.

Our special concern is the model where there is a diversity in optical properties of nanodipoles and nanocircuits. The reason for inhomogenety might then be the dispersion of geometrical scales of metaatoms which converts into a spectral inhomogeneous resonance line not necessarily symmetric. The dephasing, under the action of the first pulse, and subsequent rephrasing, caused by the second pulse, possibly results in echo effect.

This effect was reliably detected in our numerical simulations. It is established that the observed response is an oscillatory echo – the characteristic effect for an ensemble of nonlinear oscillatiors repeatedly excited by ultrashort pulses. Our preliminary calculations show that the oscillatory echo effect in the ensemble of metaatoms is stronger when the pulsation frequency falls into the NRI band.

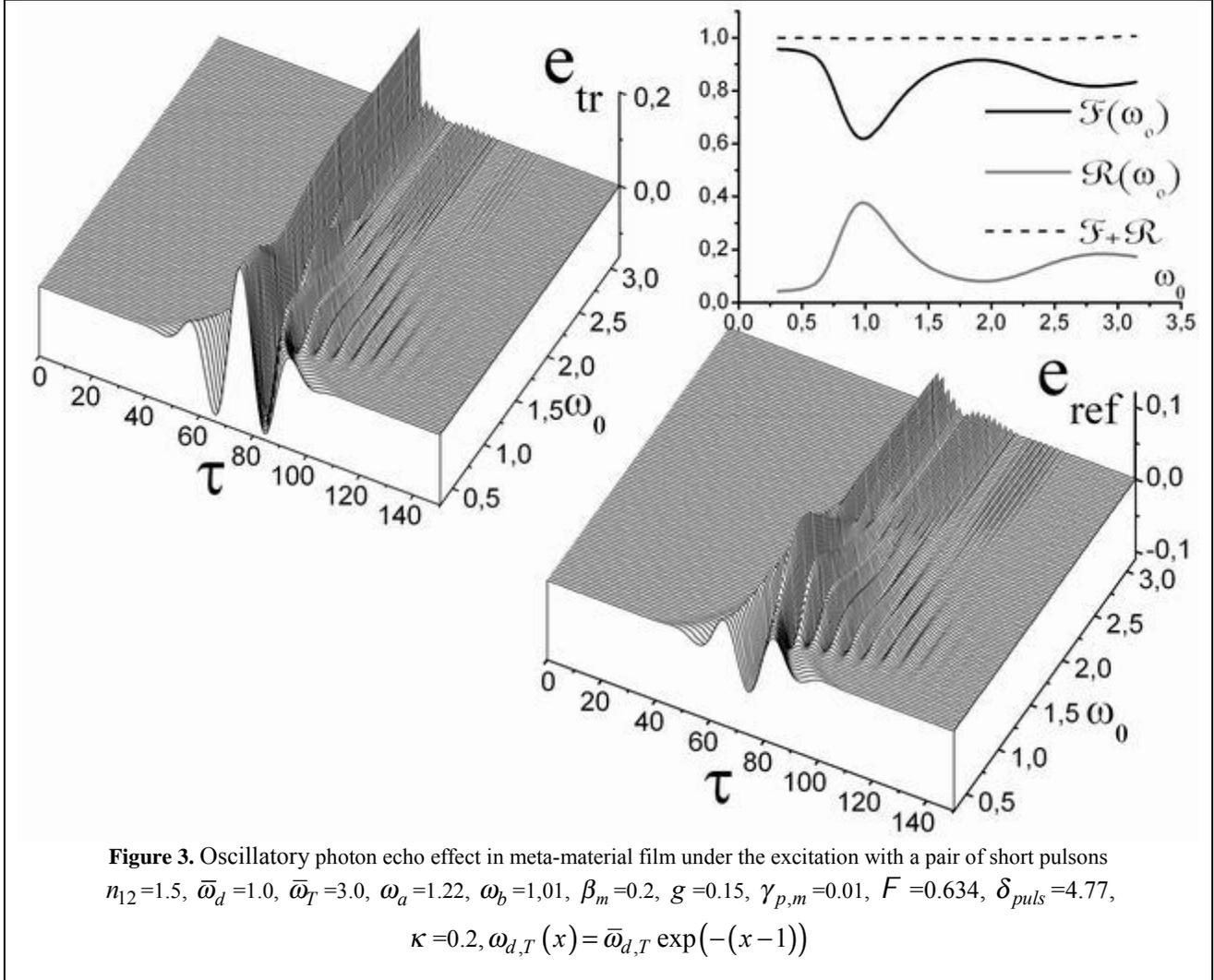

**Figure 3.** Oscillatory photon echo effect in meta-material film under the excitation with a pair of short pulsons
$n_{12}$ =1.5, $\bar{\omega}_d$ =1.0, $\bar{\omega}_T$ =3.0, $\omega_a$ =1.22, $\omega_b$ =1,01, $\beta_m$ =0.2, $g$ =0.15, $\gamma_{p,m}$ =0.01, $F$ =0.634, $\delta_{puls}$ =4.77, $\kappa$ =0.2, $\omega_{d,T}(x) = \bar{\omega}_{d,T} \exp(-(x-1))$

The transmission and reflection of a single USP is accompanied by a well-pronounced coherent effect of field nutation in the later time moments. That clearly takes place when the pulson modulation frequency is close to the resonance frequency of the model.

We believe that the further studies in this field will open the opportunities to utilize the oscillatory echo and other coherent effects as a diagnostic tool to explore metamaterials.

## ACKNOWLEDGEMENTS

We would like to express our gratitude to Askhat M. Basharov, Ildar R. Gabitov and Natalia M. Litchinister for valuable discussions and criticism on the problem under consideration. One of the authors (AIM) thanks the Department of Mathematics of University of Arizona for the support and hospitality while he looked into the problem concerned. This study was supported by Russian Foundation for Basic Research (grant № 06-02-16406).